https://doi.org/10.3847/1538-4357/ac5898

# GCM-motivated Multidimensional Temperature Parametrization Scheme for Phase-curve Retrieval

Ian Dobbs-Dixon[1,2,3] and Jasmina Blecic[1,3]
[1] Department of Physics, New York University Abu Dhabi, PO Box 129188, Abu Dhabi, UAE; idd1@nyu.edu
[2] Center for Space Science, NYUAD Institute, New York University Abu Dhabi, PO Box 129188, Abu Dhabi, UAE
[3] Center for Astro, Particle and Planetary Physics (CAP³), New York University Abu Dhabi, PO Box 129188, Abu Dhabi, UAE

## Abstract

We present a novel physically motivated, parametrized temperature model for phase-curve retrieval, able to self-consistently assess the variation in thermal structure in multidimensions. To develop this approach, we drew motivation from both full three-dimensional general circulation models and analytic formulations, accounting for the dominant dynamical feature of tidally locked planets, the planetary jet. Our formulation shows notable flexibility. It can generate planetary jets of various characteristics and redistribution efficiencies seen in the literature, including both standard eastward and unusual westward offset hotspots, as well as more exotic configurations for potential future observations. In our modeling scheme we utilize a tractable set of parameters efficient enough to enable future Bayesian analysis and, in addition to the resolved temperature structure, we return physical insights not yet derived from retrievals: the amplitude and the phase offset, and the location and the extent of the equatorial jet.

*Unified Astronomy Thesaurus concepts:* Exoplanet atmospheres (487); Exoplanet structure (495)

## 1. Introduction

The multidimensional nature of planetary atmospheres presents many challenges to modeling efforts. For decades our understanding of complex exoplanetary atmospheres has depended largely on two space telescopes, Spitzer and Hubble, and several ground-based telescopes (Knutson et al. 2012, 2014; Kreidberg et al. 2014b, 2014a; Lockwood et al. 2014; Brogi et al. 2018). Although the number of confirmed exoplanets and numerical and analytical approaches to characterize their atmospheres have grown significantly over the years, the limited wavelength coverage and spectral resolution of the telescopes have allowed us only partial insight into their envelopes (Deeg & Belmonte 2018). With some confidence we have been able to speak about global hemispheric averages and one-dimensional (1D) quantities, but the complicated three-dimensional (3D) structure has remained elusive.

With the recent James Webb Space Telescope (JWST), Transiting Exoplanet Survey Satellite (TESS), and CHaracterizing ExOPlanet Satellite (CHEOPS) launches and the expectation of two new missions in the near future, PLAnetary Transits and Oscillations of stars (PLATO) and the Atmospheric Remote-sensing Infrared Exoplanet Large-survey (ARIEL), our prospects of unveiling the multidimensional atmospheric structure are dramatically improving (Stevenson et al. 2016; Edwards et al. 2019). The number of exoplanets available for a thorough atmospheric characterization will quickly jump by an order of magnitude, yielding opportunities to select excellent targets over a wide range of properties (e.g., radius, mass, distance to the host star, stellar type).

Theoretical general circulation models (GCMs) describing exoplanetary 3D structure have become very advanced in recent years (e.g., Showman et al. 2009; Heng et al. 2011; Dobbs-Dixon & Agol 2013; Mayne et al. 2014; Rauscher & Kempton 2014; Parmentier et al. 2016; Mendonca et al. 2018; Lee et al. 2015, 2016; Drummond et al. 2018; Lines et al. 2018; Cho et al. 2021). In parallel, retrieval frameworks are getting faster with the use of high-performance computers, more advanced optimization techniques, and machine learning, allowing us to improve our assumptions (e.g., Madhusudhan & Seager 2010; Benneke & Seager 2012; Line et al. 2013; Tremblin et al. 2015; Waldmann et al. 2015; Blecic 2016; Barstow et al. 2017; Gao & Benneke 2018; Kilpatrick et al. 2018; Mollière et al. 2019; J. Blecic et al. 2022, in preparation; Barstow 2020; Cubillos & Blecic 2021). However, most of these analyses make assumptions of 1D hemispherically averaged, spherically symmetric atmospheres, and apply a single 1D $T(p)$ profile over the entire observed hemisphere. In the case of phase curves, the studies have recognized that the properties should vary with orbital phase, but have conducted retrievals on each phase independently (e.g., Stevenson et al. 2014, 2017; Venot et al. 2020).

Attempts to assess the multidimensional atmospheric structure in retrieval started in 2016, when Feng et al. (2016) explored the biases resulting from 1D assumptions within a Bayesian atmospheric retrieval. They showed that for WASP-43b synthetic HST and Spitzer phase-curve data the methane abundance was constrained to incorrect values under the 1D assumption. In Blecic et al. (2017) we explicitly compared the secondary-eclipse temperature structure produced with a 3D simulation and the retrieved 1D model. We show that the retrieved model did not exist anywhere on the planet, demonstrating in this way that tidally locked planets cannot be characterized in 1D. This idea has been furthered in forward modeling frameworks by several groups. Flowers et al. (2019) and Beltz et al. (2021) calculated very-high-resolution emission spectra from a suite of GCM models for a range of planetary rotation rates, while Caldas et al. (2019) developed a tool to generate transmission spectra from 3D temperature profiles and applied it to GJ 1214b and WASP-121b (Pluriel et al. 2020).

Based on these works, a growing number of groups are incorporating multidimensionality into retrievals. Broadly speaking, these approaches define several representative temperature-pressure profiles and combine their respective contributions using novel geometrical approaches that account for the changing viewing angle throughout the planet's orbit. In this manner, Changeat et al. (2019) defined three distinct profiles (day, night, terminator), while MacDonald et al. (2020) used varying toy models for east and west terminators to demonstrate that retrievals that assume a single temperature-pressure profile give erroneously cold temperatures at the terminators. Feng et al. (2020) extended their 2016 work to allow for viewing from arbitrary orbital phases and applied it to the phase curve of WASP-43b. Taylor et al. (2020) used an analytic formulation and toy models to explore the signal-to-noise ratio required to correctly identify an inhomogeneous temperature structure using two temperature-pressure profiles. Irwin et al. (2020) began the exploration of latitudinal temperature dependence by imposing a cosine term when calculating disk-averaged spectra. Finally, a slightly different approach is taken by the recently initiated work of K. Chubb et al. (2022, in preparation) who derive a 3D temperature model based on the diffusion equation. While this attempts to use a physical phenomenon when going from 1D to 3D, there may be limitations associated with this approach as horizontal advection cannot be fully modeled as a diffusive process.

Though these approaches recognize the multidimensional nature of the atmosphere, none of them start from the intrinsic atmospheric processes and address the fundamental dynamical feature of close-in tidally locked planets, the circumplanetary equatorial jet. Short-period planets (from hot Jupiters to terrestrial planets), presumably tidally locked and highly irradiated, tend to exhibit strong dynamical forcing (e.g., Showman & Guillot 2002; Dobbs-Dixon & Lin 2008). The uneven and intense stellar irradiation of their atmospheres leads to large longitudinal temperature gradients closely tied to supersonic equatorial and mid-latitude jets that advect energy around the planet. Inherently 3D, complicated atmospheric dynamics of these planets leads to significant compositional and temperature variations, most prominently seen with changes in longitude. To adequately constrain a plausible atmospheric structure, one must rigorously compare our understanding of these physical and chemical processes with the observational constraints. Thus, combining our theoretical knowledge gathered from GCMs with a robust statistical, observationally driven retrieval approach, utilizing photometric and spectroscopic phase-curve observations, is our best way forward to assess the complicated 3D atmospheric structure of exoplanetary objects.

Phase-resolved emission spectrometry (phase-curve observations) carries the most comprehensive information about planetary envelopes, atmospheric dynamics, chemical processes, and radiative energy balance (e.g., Knutson et al. 2007; Cowan & Agol 2008; Stevenson et al. 2014; Venot et al. 2020). These observations allow us to map the entire surface of a planet, as for a tidally locked planet the orbital phase is equivalent to its rotational phase, providing, in addition to the altitudinal information, the longitudinal and latitudinal constraints on the atmospheric composition, thermal structure, and energy distribution. Furthermore, in the thermal infrared, the amplitude of the phase variation determines the day–night temperature contrast and the phase offset reveals the longitude of the planet's hottest point. These dynamical features derive primarily from the equatorial jet redistributing heat from the hot dayside to the cooler nightside.

In this paper we use analytic formulations and GCM models to account for the planetary jet, and fundamentally link all orbital phases in a more physically consistent way. Our parameterization scheme accounts for the energy redistribution around the planet by including both advective and radiative timescales. Here, we present a forward model and we explore the flexibility of our formulation. In a following paper, we will present the implementation of this model in retrieval, utilizing phase-curve observations. Such application will allow us to self-consistently tie together temperature profiles across all longitudes and latitudes, enabling a simultaneous retrieval of spectra at all orbital phases, and returning important constraints on the atmospheric dynamics.

The paper is structured as follows. In Section 2 we briefly outline our methodology. In Section 3 we define our 2D temperature model, list the model parameters, and explore the versatility of our assumptions. In Section 4 we describe additional assumptions for the full 3D representation, and in Section 5 we summarize our results.

## 2. Methodology

We use radiative-hydrodynamic (RHD) simulations, employing a double-gray radiative scheme to model both the radiative and the advective contributions to the total energy budget. The radiative portion of our RHD model is described in Dobbs-Dixon & Agol (2013).

The radiative contribution was chosen to match the analytic formulation of Guillot (2010), while to isolate and parameterize the advective contribution we explicitly use the RHD outputs (see Sections 3.1 and 3.2). Our prescription models all the crucial features seen in the GCM simulations responsible for the advection of energy from the day- to the nightside of the planet. By using an amenable set of parameters, we allow for a large flexibility in our models, covering both solutions seen in the GCM simulations and physical scenarios not yet explained by the theory but observed among exoplanets (e.g., Dang et al. 2018; von Essen et al. 2020). Our temperature parameterization is physically motivated and fully analytic and it allows us to calculate the planetary 3D temperature structure in microseconds, as opposed to days or weeks using standard GCM simulations. A reasonable set of model parameters also permits for an efficient Bayesian analysis, allowing for thousands of models to be explored in a short period of time. Such formalism returns the physically consistent resolved temperature structure, but also, for the first time, the fundamental properties of the jet.

## 3. 2D Temperature Profile

To develop our 2D ($p$, $long$) temperature parameterization scheme we start with a static, plane-parallel, isotropically scattered atmosphere in local thermodynamic equilibrium, assuming the condition for a purely radiative equilibrium:

$$\int_0^\infty \kappa_\nu (J_\nu - B_\nu) d\nu = 0. \tag{1}$$

Here, $\kappa_\nu$ is the absorption coefficient, $J_\nu$ is the source function, and $B_\nu$ is the Planck function. To derive the temperature structure, we follow Guillot (2010) and introduce two Eddington coefficients, $f_{kth}$ and $f_{hth}$, for the thermal radiation pressure and flux, respectively (Eddington 1916). We decouple the thermal and visible radiation and parameterize the opacities using the mean visible and thermal opacities, defining their ratio as $\gamma = \kappa_v/\kappa_{th}$. The gas temperature as a function of the optical depth, $\tau$, can then be written as

$$
\begin{aligned}
T_{rad}^4(\tau) = {} & \frac{3T_{int}^4}{4}\left[\frac{1}{3 f_{kth}} + \frac{\tau}{3 f_{kth}}\right] \\
& + \frac{3T_{irr}^4}{4}\mu\left[\frac{1}{3 f_{hth}} + \frac{\mu}{3\,\gamma\, f_{kth}}\right] \\
& + \frac{3T_{irr}^4}{4}\mu\left[\left(\frac{\gamma}{3\,\mu} - \frac{\mu}{3\,\gamma\, f_{kth}}\right)e^{-\gamma\,\tau/\mu}\right].
\end{aligned}
\tag{2}
$$

The first term in this equation describes the spherically symmetric interior energy contribution, where $T_{int}$ is the internal temperature of the planet, $\tau$ is the infrared optical depth, given as $\tau = p\,\kappa_{th}/g$, $p$ is the pressure, $g$ is the planetary gravity, and $\mu = \cos(long) * \cos(lat)$, with "$long$" being the longitude and "$lat$" being the latitude. The second and third terms define the angle-dependent irradiative contribution, where $T_{irr}$ is the irradiation temperature, given as

$$
T_{irr} = T_*\left(\frac{R_*}{a}\right)^{1/2},
\tag{3}
$$

with $T_*$, $R_*$, and $a$ being the stellar surface temperature, radius, and semimajor axis, respectively.

Equation (2) is easily solved analytically, and we denote this radiative temperature contribution as $T_{rad}^4$. However, this purely radiative approach does not account for the important advection of thermal energy by the jet. The planetary jet links temperature profiles across all planetary phases by removing energy from the dayside and adding energy onto the nightside.

At a fundamental level, one can introduce an advective flux by modifying the radiative equilibrium condition to be

$$
\int_0^\infty \kappa_\nu(J_\nu - B_\nu)\,d\nu = q\nabla T,
\tag{4}
$$

where $q\,\nabla\,T$ represents the total advective flux. Total radiative absorption and emission must now be additionally balanced with thermal energy advected in and out of an area. The vertical temperature profile can then be written as

$$
\begin{aligned}
T_p^4(\tau) = {} & \frac{3T_{int}^4}{4}\left[\frac{1}{3 f_{hth}} + \frac{\tau}{3 f_{kth}}\right] \\
& + \frac{3T_{irr}^4}{4}\mu\left[\frac{1}{3 f_{hth}} + \frac{\mu}{3\,\gamma\, f_{kth}}\right] \\
& + \frac{3T_{irr}^4}{4}\mu\left[\left(\frac{\gamma}{3\,\mu} - \frac{\mu}{3\,\gamma\, f_{kth}}\right)e^{-\gamma\,\tau/\mu}\right] \\
& - \frac{\pi}{\sigma}\left\{\left(\frac{1}{f_{hth}} + \frac{\tau}{f_{kth}}\right)\int_0^\infty q\nabla T dm\right. \\
& \left. + \frac{\tau}{f_{kth}}\int_0^\infty\left(\frac{m'}{m} - 1\right)q\nabla T\,dm' - q\nabla T\right\}.
\end{aligned}
\tag{5}
$$

The last two terms, coming from the advective sources and sinks in the atmosphere, significantly complicate the analytic solution of this equation. Here, we simply denote each of the terms as

$$
T_p^4 = T_{rad}^4 + T_{adv}^4,
\tag{6}
$$

where $T_{rad}^4$ carries both internal and irradiation terms (first, second, and third term in Equation (2)) and $T_{adv}^4$ carries the last two advective terms in Equation (5) (the $\tau$ dependence is dropped for clarity). Several other authors (e.g., Guillot 2010; Hansen 2008; Burrows et al. 2008) have formulated similar expressions to Equation (5), but a purely analytic solution to the total temperature structure has so far proven elusive. Thus, to formulate an expression for $T_{adv}^4$ we have turned to results from our GCM simulations. We note that $T_{rad}^4$ and $T_{adv}^4$ are not truly decoupled, as several parameters ($\gamma$, $f_{hth}$, and $f_{kth}$) appear in both terms of Equation (6). However, since the model was formulated for phase-curve retrieval, it will reveal the global planetary properties (the consistent 3D temperature structure and jet characteristics) from the coupled, total planetary temperature ($T_p$) model (not the individual components), making irrelevant (or secondary) the fact that the individual components are in fact coupled.

### 3.1. Analytic Radiative Component

To parameterize individual contributions to $T_p$ from Equation (6), we simulate a planet resembling HD 189733b, utilizing our RHD model with a dual-gray radiative scheme. We utilize two sets of GCM simulations: one fully self-consistent, cloud-free model and the other identical except for the introduction of a very large artificial damping of the velocities. By damping the velocities to near zero, we are effectively shutting off any advective contribution. The resulting solution is purely radiative (and we denote it as $T_{rad}$) and agrees quite well with the analytic solution from Guillot (2010; see Figures 1 to 3 and Section 3.2) and the expression given in Equation (2), which is fully described by five parameters: $T_{int}$, $T_{irr}$, $\gamma$, $f_{hth}$, and $f_{kth}$ (refer to Guillot 2010 and Blecic et al. 2017 for an exhaustive study of the effect of these parameters on the temperature-pressure profiles). We note that in our parameterization the Eddington parameters ($f_{hth}$ and $f_{kth}$) can be effectively excluded from the model. As Guillot (2010) states in their Section 2.4, Eddington coefficients can carry only a few possible values; however, only the solution assuming an isotropic radiation field ($f_{hth} = \frac{1}{3}$, $f_{kth} = \frac{1}{2}$) closely matches the theoretical solution at higher pressures. In our subsequent analysis, we adopt these values, but we also leave the user an option to use different values. This leaves us with only three free parameters for the radiative solution ($T_{int}$, $T_{irr}$, $\gamma$).

### 3.2. Parameterized Advective Component

By subtracting the GCM's radiative solution, $T_{rad}^4$, from the full GCM solution, we are able to isolate the advective contribution, $T_{adv}^4$, and subsequently model it. In this section, we present our parameterization of the advective component in 2D, as a function of pressure and longitude. In Section 4, we introduce additional parameters for a full 3D representation (as a function of pressure, longitude, and latitude).

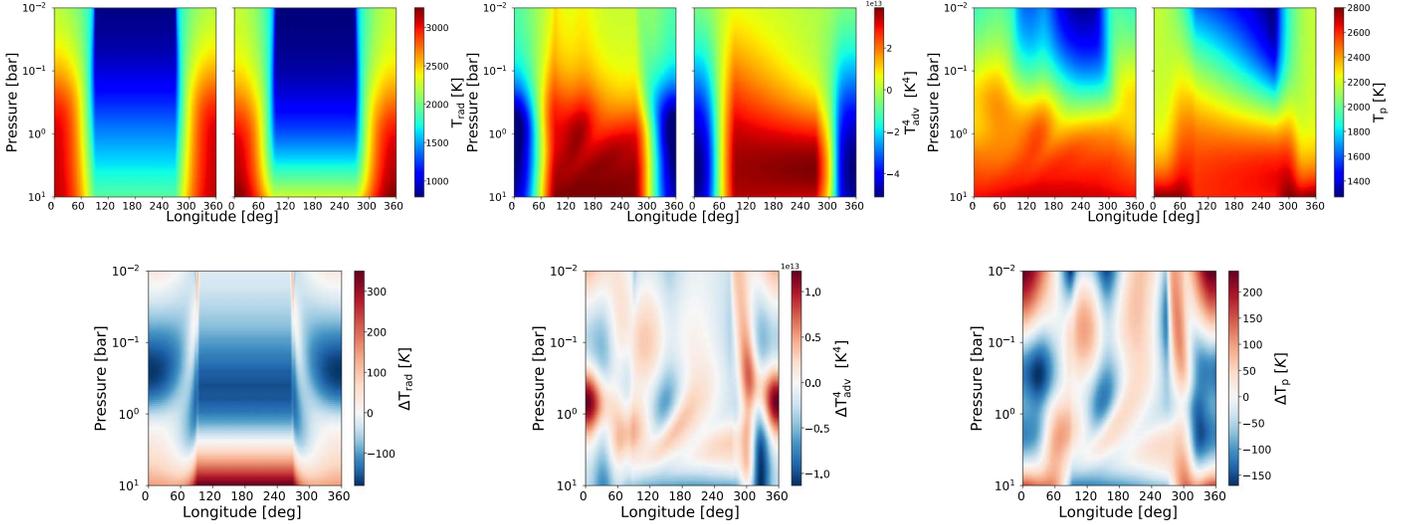

**Figure 1.** Hot case, with an irradiation temperature of 2500 K. Top panels: equatorial temperature distributions. From left to right: (1) Radiative component: strongly damped dual-gray HD 189733b GCM simulation, $T_{\rm rad}$ (left) and analytic best-matched radiative solution (right), as a function of longitude and pressure. The substellar point is located at 0° longitude. (2) Advective component: dual-gray 189733b GCM solution, $T_{\rm adv}^4$, (left) calculated by subtracting the damped (panel 1, left) simulation from the full (panel 3, left) GCM simulation and our best-matched parametrized advective model (right). Negative regions on the dayside imply advection acts to extract energy from this region, while positive areas denote deposition of energy by the jet. (3) Total planetary temperature: full GCM simulation, $T_{\rm p}$, (left) and the best-matched parametrized total planetary temperature model (right). Bottom panels: temperature difference between the GCM simulations and the parametrized models corresponding to the panels on the top.

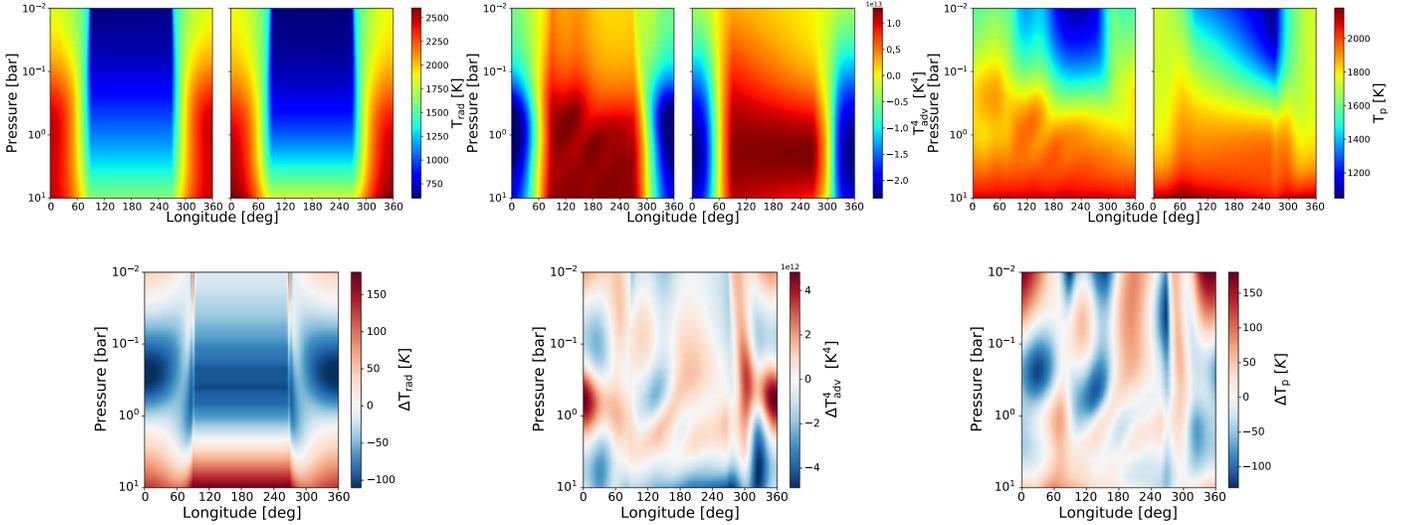

**Figure 2.** Medium case, with an irradiation temperature of 2000 K. Top and bottom panels: same as in Figure 1.

To derive our parameterization, we ran three different GCM simulations, a hot, medium, and cold case, assuming irradiation temperatures of 2500, 2000, and 1500 K, respectively, and carefully studied all the features seen in the results. Figures 1 to 3 show a comparison between the radiative, advective, and planetary temperatures generated using these GCM simulations and our parametrized model. We also show the temperature differences for each component, calculated by subtracting the GCM results from the most closely matching parametrized model.

For the radiative component ($T_{\rm rad}$; Figures 1, 2 and 3, left double panel), in both the damped GCM simulations (left sides) and our parametrized solutions (right sides), as expected, the profiles are getting colder going from the substellar point toward the nightside. On the nightside, the temperature of the radiative profiles remains constant, as there is no contribution from the stellar flux and no mechanism to transfer the energy to the nightside of the planet. We see nice matches between the GCM simulations and the parametrized models in all three cases.

For the advective component ($T_{\rm adv}^4$; Figures 1, 2 and 3, middle double panels), we see that for each case the temperatures are negative on the dayside and positive on the nightside, illustrating the extraction of energy from the dayside and deposition on the nightside. The left side panels, generated by subtracting the strongly damped GCM simulation from the full GCM simulation, clearly show that the efficiency of 2D advection is a strong function of both longitude and pressure. Therefore, to parameterize $T_{\rm adv}^4$ and allow for sufficient flexibility in the parameters, such that one can model a wide range of jet characteristics, we identify a number of different characteristics evident in the figures. First, in longitude, we distinguish three segments across the planetary surface: east dayside ($seg_1$, 0°–90°), nightside ($seg_2$, 90°–270°), and west dayside ($seg_3$, 270°–360°). In each of these segments $T_{\rm adv}^4$ has a different functional dependence in longitude (see next paragraph). Second, to account for the pressure dependence, also clearly seen in the figures, the magnitude of this term within the segments is allowed to vary. In particular, we find that the values of $T_{\rm adv}^4$ at the *interfaces* between the segments (three in total, approximately corresponding to 0°, 90°, and 270° longitude) are well modeled as Gaussians in log pressure. At each interface, we introduce parameters for the amplitude ($A_i$, in units of K[4]), width ($\sigma_i$, in units of bar), and center ($c_i$, in units of bar) of these Gaussians, which can freely vary. We further note that the cores of equatorial jets found in GCM simulations of tidally locked planets are usually isobaric across the planet, allowing us to define only a single central pressure for all three interfaces (e.g., Rauscher & Menou 2010). This leaves us with a total of seven parameters ($A_1$, $A_2$, $A_3$, $\sigma_1$, $\sigma_2$, $\sigma_3$, $c$). Utilizing these, the value of $T_{\rm adv}^4$ at each of the three interfaces, denoted by $\Gamma(p)$, can be calculated as

$$\Gamma_i(p) = A_i \exp\left(-\left[\frac{\log(p) - \log(c)}{\log(\sigma_i)}\right]^2\right). \tag{7}$$

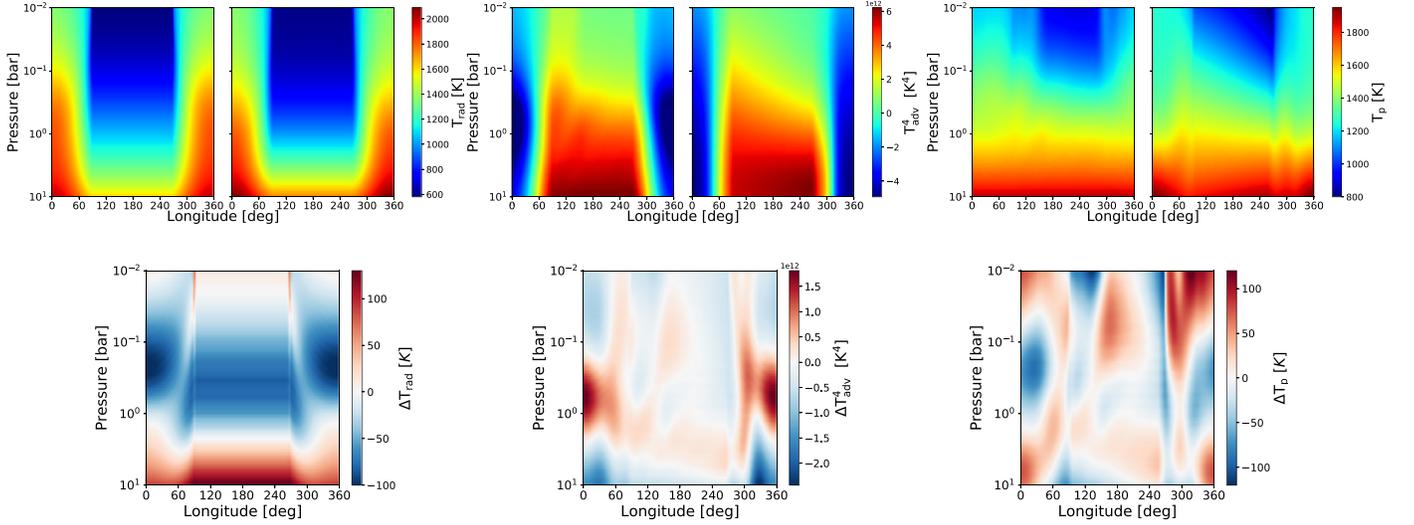

**Figure 3.** Cold case, with an irradiation temperature of 1500 K. Top and bottom panels: same as in Figure 1.

Here, $p$ is the pressure and the index-$i$ denotes the interfaces: (i) between east and west daysides; (ii) between east dayside and the nightside; and (iii) between the nightside and west dayside. Figure 4 shows an example of the Gaussian $T_{\rm adv}^4$ profiles for a set of fiducial values.

To parameterize the behavior of $T_{\rm adv}^4$ across the three regions in longitude ($seg_1$, $seg_2$, $seg_3$) we again turn to the GCM results and find that the east and west dayside regions are well modeled with the sum of parabolic and error functions, while the nightside is best fit with a linear function (Figure 5). We further introduce longitudinal offset parameters ($\phi_2$ and $\phi_3$, in units of degrees), that can shift the two interfaces (dotted vertical lines in Figure 5, right panel) in any direction, determining in this way where most of the energy transferred from the dayside will be deposited. This induces eastward or westward shifts of the temperature profiles, suggestive of the observed phase offsets (see also Section 3.2.2).

Thus, we model each of these segments analytically as

$$seg_1(p, \varphi) = \frac{E_{\rm parab}(p, \varphi) + E_{\rm erf}(p, \varphi)}{2}, \quad \in[\varphi_1, \varphi_2]$$

$$seg_2(p, \varphi) = lin(p, \varphi), \quad \in[\varphi_2, \varphi_3]$$

$$seg_3(p, \varphi) = \frac{W_{\rm parab}(p, \varphi) + W_{\rm erf}(p, \varphi)}{2}, \quad \in[\varphi_2, \varphi_3], \tag{8}$$

where $\varphi_1 = 0°$, $\varphi_2 = 90° \pm \phi_2$, $\varphi_3 = 270° \pm \phi_3$. The parabolic, linear, and error functions are given as

$$E_{\rm parab}(p, \varphi) = \frac{\Gamma_2(p) - \Gamma_1(p)}{(\varphi_2 - \varphi_1)^2}(\varphi - \varphi_1)^2 + \Gamma_1(p) \tag{9}$$

$$W_{\rm parab}(p, \varphi) = \frac{\Gamma_3(p) - \Gamma_1(p)}{(\varphi_3 - 2\pi - \varphi_1)^2}(\varphi - 2\pi - \varphi_1)^2 + \Gamma_1(p) \tag{10}$$

$$lin(p, \varphi) = \frac{\Gamma_3(p) - \Gamma_2(p)}{\varphi_3 - \varphi_2}(\varphi - \varphi_3) + \Gamma_3(p), \tag{11}$$

$$E_{\rm erf}(p, \varphi) = \frac{\Gamma_2(p) + \Gamma_1(p)}{2} + \frac{\Gamma_2(p) - \Gamma_1(p)}{2}$$
$$\times \operatorname{erf}\left[\frac{4}{\varphi_2 - \varphi_1}(\varphi - \frac{\varphi_2 + \varphi_1}{2})\right], \tag{12}$$

$$W_{\rm erf}(p, \varphi) = \frac{\Gamma_3(p) + \Gamma_1(p)}{2} + \frac{\Gamma_3(p) - \Gamma_1(p)}{2}$$
$$\times \operatorname{erf}\left[\frac{6}{\varphi_3 - \varphi_1 - 2\pi}(\varphi - \frac{2\pi + \varphi_2 + \varphi_1}{2})\right]. \tag{13}$$

In the above equations, $\Gamma_i(p)$ is defined by Equation (7), $p$ is pressure, and $\varphi$ is longitude. Equations (9) and (10) denote the eastern and western parabolic expressions, Equation (11) gives the linear expression for the nightside change in longitude, while Equations (12) and 13 denote eastern and western error functions.

Figure 5 shows a comparison between the GCM advective component for the medium case (Figure 2, panel 2) at a pressure of ~0.1 bar, and the corresponding parametrized

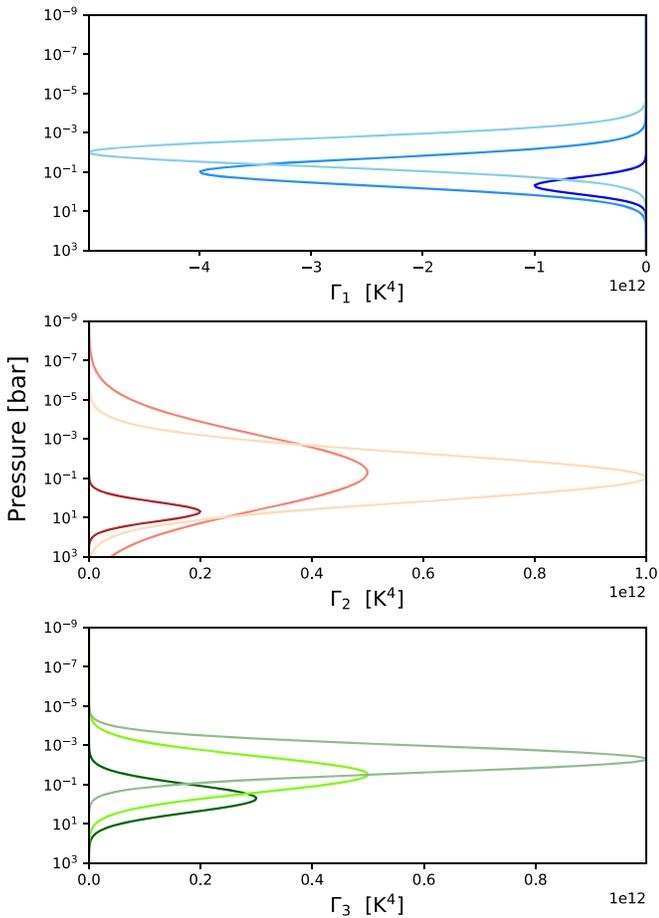

**Figure 4.** Left: Gaussian pressure parameters for different values of amplitude, width, and center (Equation (7)). The $\Gamma_1$ parameter describes the transfer of energy between $\sim$0 and $\sim$90°, $\Gamma_2$ between $\sim$90° and $\sim$270°, and $\Gamma_3$ between $\sim$270° and $\sim$360° longitude.

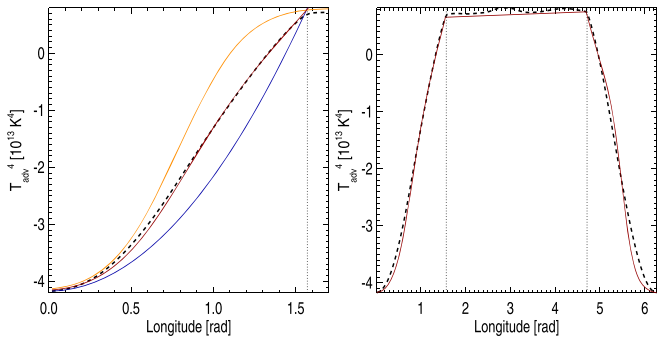

**Figure 5.** Comparison between the parametrized model and $T^4_{\mathrm{adv}}$ derived from the GCM (Figure 2, panel 2). Plots are shown for a pressure of $\sim$0.1 bar for a fiducial set of parameters. Left: contributions from both the error (orange) and parabolic (blue) functions to the overall parameterized model (red) of $seg_1$. The thick dashed black line is the result from the GCM. Right: comparison which spans all longitudes and shows a good match between the model and simulation for all three segments. Dashed vertical lines denote the intersection between the segments, where Equation (7) is valid.

model generated using Equations (8)–(13). Figure 6 shows each of the segments in a different color for all pressures (given as hue gradient) and longitudes, for the same set of model parameters as in Figure 2, panel 2, right side. For illustration purposes, we have set the location of the interfaces to be 0°,

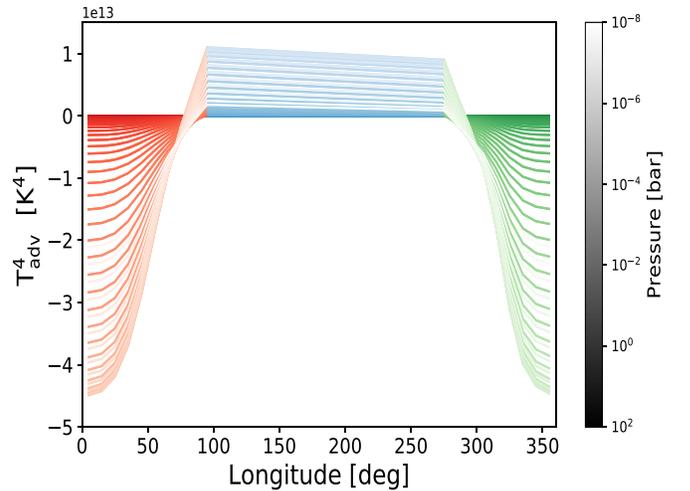

**Figure 6.** The advective temperature, $T^4_{\mathrm{adv}}$, as a function of planetary longitude for a range of pressures (Figure 2, right side of panel 2). Different colors depict the three different segments given in Equation (8). The gray gradient on the pressure color bar corresponds to each segments' line color gradient.

90°, and 270° on both Figures 5 and 6, though as shown in the above equations these can vary.

Taken together, we can now calculate a fully parameterized $T^4_{\mathrm{adv}}$ as a function of longitude and pressure. These parameterized quantities are shown in Figures 1, 2, and 3, panels 2, right sides and, as seen, they agree quite well with the GCM results shown in the same figures, panels 2, left sides.

Adding the advective term ($T^4_{\mathrm{adv}}$) to the radiative term ($T^4_{\mathrm{rad}}$), which is calculated analytically using Equation (2), we get the total 2D temperature structure for each case. Figures 1, 2, and 3, panels 3, show the comparison between the full GCM simulations and the parametrized total planetary temperature model including both radiative and advective components. As we can see, the GCM simulations agree nicely with our best-matched parametrized planetary temperature solution.

The bottom panels in Figures 1, 2, and 3, showing the temperature differences between the GCM simulations and parametrized models, demonstrate that the differences are between 5% and 10%, on the order of the uncertainties usually seen when retrieving planetary temperatures. Differences in the radiative component are most pronounced at the terminators and at depth. The differences at the terminators are due to the abrupt change in longitude in the analytic formulation that is smoothed out in the highly damped GCM simulations, while the differences in the depth are likely due to the fact that the GCM has not converged at high pressures. For the advective contribution and total planetary temperature, the differences seen in the plots are indicative of wave-like structures known to be present in GCM models, but not included in the analytic formulation.

### 3.2.1. Model Flexibility

Our analytic temperature parameterization allows us to generate physically consistent $T(p, long)$ profiles throughout the entire planet, subject to a set of three parameters governing the radiative solution, $T_{\mathrm{rad}}$ ($\kappa_v$, $\kappa_{\mathrm{th}}$, $T_{\mathrm{int}}$; see Equation (2)), and nine governing the advective solution, $T_{\mathrm{adv}}$ ($A_1$, $A_2$, $A_3$, $\sigma_1$, $\sigma_2$, $\sigma_3$, $c$, $\phi_2$, $\phi_3$). The flexibility of our parameterization, also seen in Figures 1–3, allows one to explore a wide range of jet parameters and characteristics, crucial for use in retrieval.

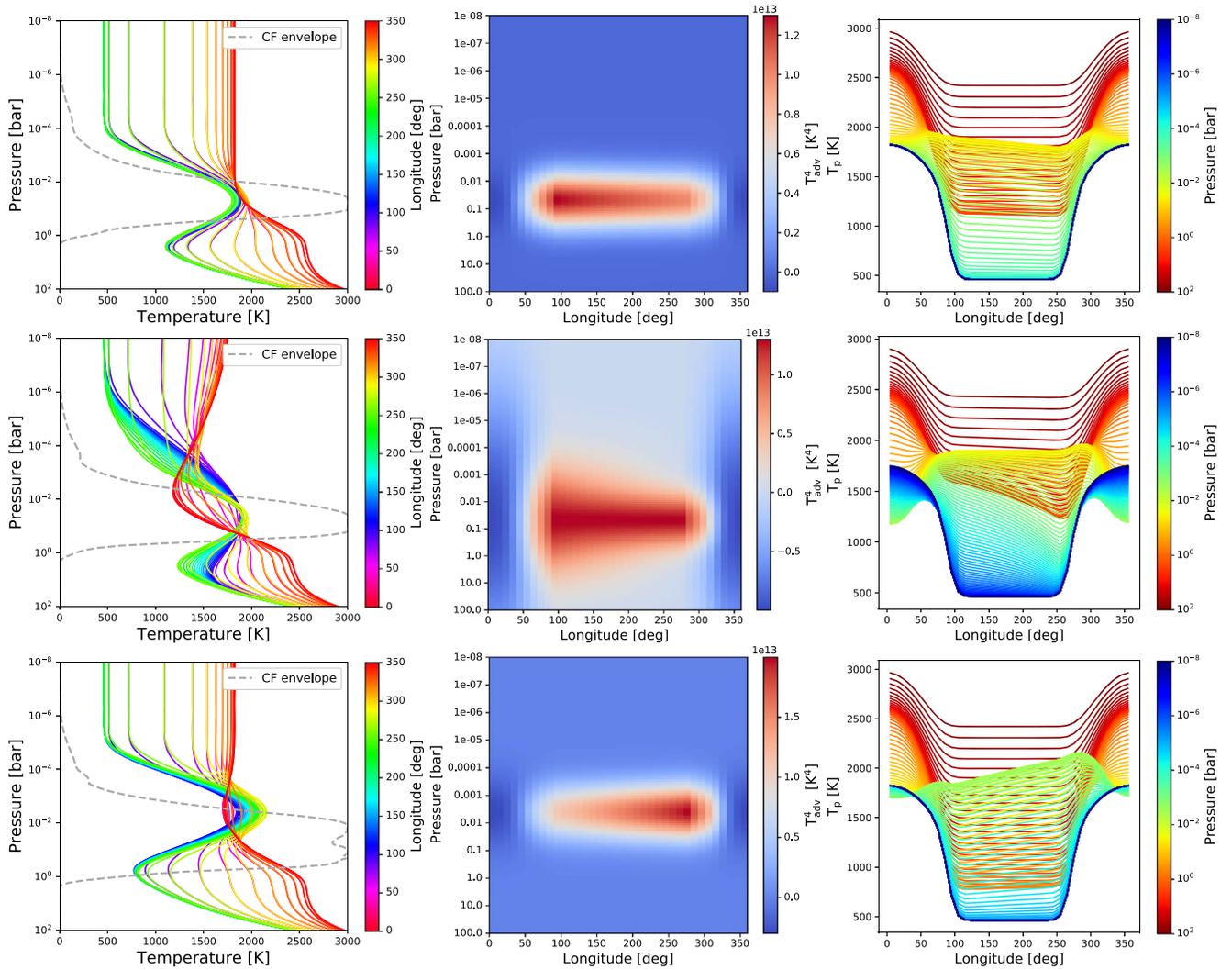

**Figure 7.** Three examples of our parameterization scheme, demonstrating the flexibility of our approach. Specific parameters are listed in Table 1. Left panels: temperature profiles at the equator across all longitudes, with the contribution function envelope (given in dashed gray) calculated at 4.5 $\mu$m. Middle panels: $T_{adv}^4$ at the equator. Right panels: corresponding total planetary temperatures as a function of longitude and pressure.

Figure 7 shows several temperature structures generated using our prescription, demonstrating the effectiveness of our assumptions (model parameters are listed in Table 1). The left column shows temperature profiles at the equator across all longitudes, corresponding to the different jets ($T_{adv}^4$) shown in the middle column, together with the global planetary temperatures across all pressure shown in the right column. As presented, we can decide on the location of the jet, whether the energy from the dayside will be taken from wider or narrower pressure range, and on which longitudes and pressures most of this energy will be deposited. The differences in the radial extent, strength, and direction of the jets are also clearly visible when comparing the three middle figures. This allows us to explore the standard eastward temperature offsets, but also unusual configurations such as a westward offset (e.g., Corot-2b, WASP-33b, as in Dang et al. 2018 and von Essen et al. 2020, respectively).

### 3.2.2. Flux, Phase Curves, Spectra

Once we have specified the temperature structure, we can easily generate flux for each wavelength and at each point on the planetary surface. To calculate intensities, we perform radiative-transfer calculations using the open-source PYRATBAY framework for exoplanet atmospheric modeling, spectral synthesis, and Bayesian retrieval (Cubillos & Blecic 2021). In Figures 8 and 9, we use our 2D temperature parameterization scheme and assume hydrogen-dominated atmospheres, accounting for major spectroscopically active species and their opacities (CO, $CO_2$, $CH_4$, $H_2O$, HCN, $NH_3$). Our code can, however, assume any chemical composition and account for any opacity sources from ExoMol[4] and HITRAN/HITEMP[5], as well as other databases (see Sections 2.3.1–2.3.5 in Cubillos & Blecic 2021).

To calculate flux, we first divide the atmosphere into equally spaced longitude–latitude boxes (visible pixels on Figures 8 and 10), where each box has its own temperature profile, calculated either using our 2D temperature parameterization scheme (Section 3), or the 3D scheme (Section 4). Then, we calculate the observer's projected area (Seager 2010) at each phase (Figure 10 shows examples of the observer's projected area at phases 0°s and 270°). Finally, using PYRATBAY's radiative-transfer routines, we calculate the intensity per box, convolve it with the corresponding observer's box area to get the "flux per box", generating in this way the total spatially dependent emission flux across the observable hemisphere. Repeating this routine for each phase allows us to create a phase curve for the desired wavelength. The phase curve then returns three fundamental properties of the jet: the phase offset, amplitude, and pressure extent.

**Table 1**
Figure 7 Model Parameters

| Components | $T_{rad}^4$ [K$^4$] | | | | | $T_{adv}^4$ [K$^4$] | | | | | | |
|---|---|---|---|---|---|---|---|---|---|---|---|---|
| Parameters [units] | $\kappa_v$ [$\frac{cm^2}{g}$] | $\kappa_{th}$ [$\frac{cm^2}{g}$] | $T_{int}$ [K] | $\phi_2$ [deg] | $\phi_3$ [deg] | $A_1$ [K$^4$] | $A_2$ [K$^4$] | $A_3$ [K$^4$] | $\sigma_1$ [bar] | $\sigma_2$ [bar] | $\sigma_3$ [bar] | $c$ [bar] |
| Top row | $4.0 \times 10^{-3}$ | $10^{-2}$ | 550 | 0 | 0 | $-10^{12}$ | $1.3 \times 10^{13}$ | $10^{13}$ | $10^{-5}$ | $10^{-5}$ | $10^{-5}$ | 0.05 |
| Middle row | $4.0 \times 10^{-3}$ | $10^{-2}$ | 550 | 0 | 0 | $-10^{13}$ | $1.3 \times 10^{13}$ | $1.3 \times 10^{13}$ | 0.1 | $10^{-4}$ | $10^{-5}$ | 0.05 |
| Bottom row | $4.0 \times 10^{-3}$ | $10^{-2}$ | 550 | 0 | 0 | $-3.0 \times 10^{12}$ | $1.1 \times 10^{13}$ | $2.0 \times 10^{13}$ | $2.0 \times 10^{-5}$ | $10^{-5}$ | $10^{-5}$ | 0.004 |

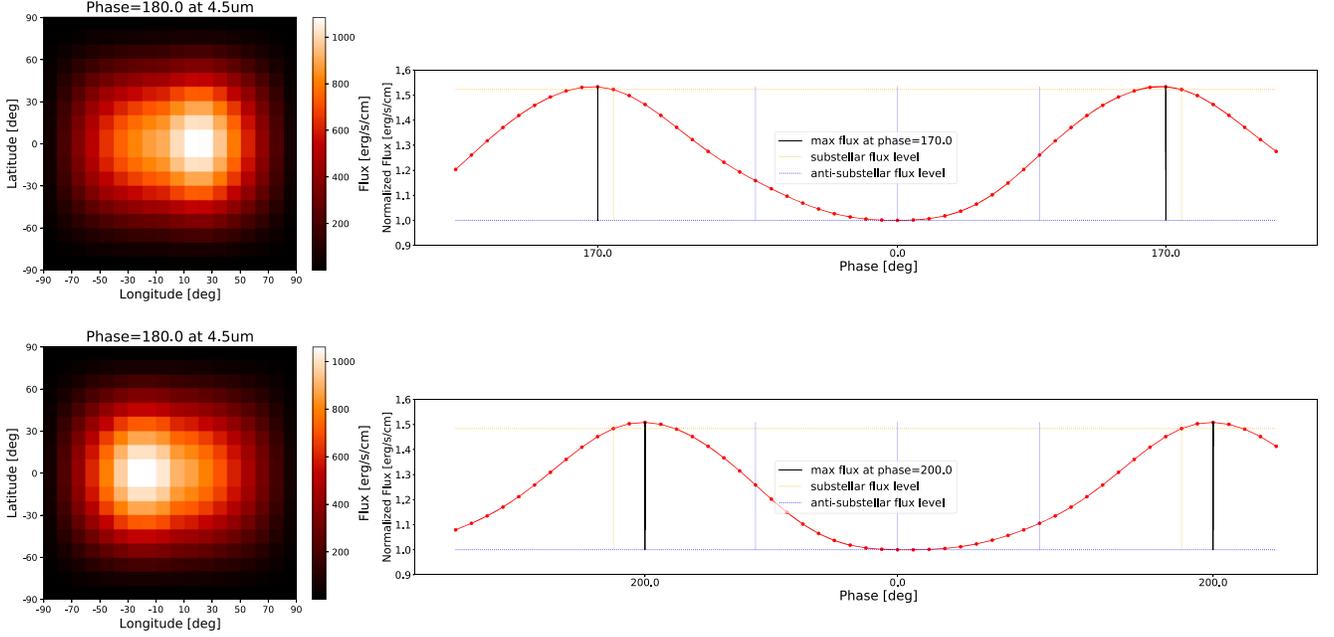

**Figure 8.** Examples of eastward and westward offsets. Left panels: spatially dependent dayside emission flux of a parameterized planet in the Spitzer 4.5 μm bandpass with visible eastward (top panel) and westward (bottom panel) offset hotspots. Right panels: the corresponding phase curves with visible phase offsets. Blue vertical lines denote the 270°.0, 0°.0, 90°.0 phases, orange vertical lines denote the 180°.0 phase, while black vertical lines denote the phase with maximum flux. The dotted orange and blue horizontal lines denote the substellar flux level (at phase 180°.0) and the antisubstellar flux level (at phase 0°.0), respectively.

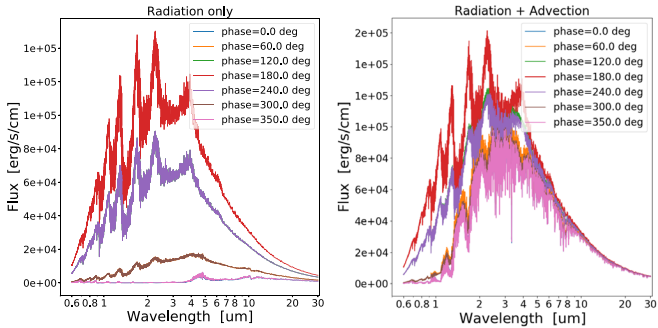

**Figure 9.** Emission spectra as a function of phase without (left) and with (right) the advective contribution (top example in Figure 7), with obvious increase in the nightside flux.

Figure 8 shows the spatially dependent emission flux in examples of the eastward and westward hotspots, and the corresponding phase curves with the phase shifts at 170° and 200°, respectively (for observers this corresponds to the peaks that occur at +10°, top panel, and −20°, bottom panel, before and after opposition, respectively). These offsets are produced for a planet observed at 4.5 μm at secondary eclipse. The pressure location of the jet can be recovered from the center parameter, $c$, while the amplitude and the phase offset can be easily extracted from the phase curve. The flexibility of our approach allows us to generate the offset hotspot at any longitude on the dayside (or even nightside) of the planet.

To generate spectra at each phase across a range of wavelengths, for each wavelength we calculate "flux per box" and add them over the observable atmosphere to get the surface flux. Figure 9 shows the phase-dependent spectra from a planet with and without an advective contribution. The increase in the flux from the nightside due to the jet's influence is clearly seen on the right panel.

## 4. 3D Planetary Structure

Our ultimate goal is to develop a physically motivated 3D $T$ ($p$, $long$, $lat$) approach for phase-curve retrieval to model the true multidimensional nature of planetary atmospheres. This will allow us to link both the planetary longitudes and latitudes in a physically consistent way.

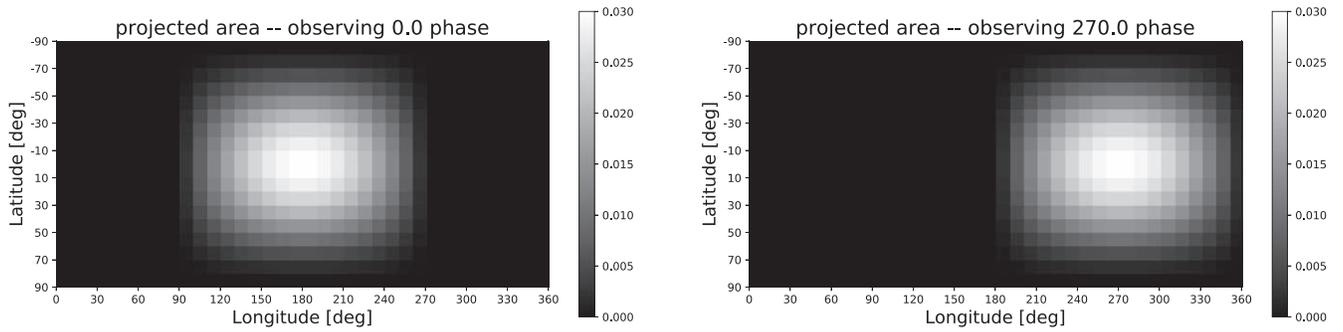

**Figure 10.** Observers projected area on the planet at orbital phases of 0° (left) when looking directly at the nightside, and 270° (right). The total hemispherical area sums to one.

**Figure 11.** Examples of advective temperature structures at a pressure of 0.02 bar, generated utilizing our 3D $T(p, long, lat)$ model and the jet characteristics shown in the top panels of Figure 7. Left: $T_{adv}^4$ with a Gaussian width of 30° in the latitudinal direction. Right: same as left panel with a Gaussian width of 55°. As seen, the advection shape, direction, and strength seen in Figure 7 are well preserved in both left and right panels.

To extend our 2D $T(p, long)$ model to 3D $T(p, long, lat)$, we again utilize the results of GCM simulations. The equatorial jet seen in tidally locked atmospheres has a latitudinal extent closely connected to the Rhines scale (Showman & Guillot 2002), thus dependent on the wind velocity and rotation rate. Fast winds and slow rotation rates (compared to Jupiter) imply wide jets. Simulations indicate that the jet usually extends between ∼±25° latitude around the equator (e.g., Dobbs-Dixon & Agol 2013). To take advantage of the north/south symmetry and mimic this behavior, we assume another Gaussian with a variable width in latitudinal direction ($\sigma_{lat}$) symmetric around the equator, and integrate it into our parameterized $T_{adv}^4$ model. The radiative part of our $T(p, long, lat)$ model will also be attenuated in latitude by including $\cos(lat)$ in the $\mu$ parameter, following Guillot (2010).

Since we are already dividing the planet into longitudinal–latitudinal cells, this allows a smooth transition to 3D when calculating flux, spectra, or phase curves. In Figure 11 we show examples of 3D advective temperature structures, for two different latitudinal extents of the jet, while in Figure 12 we show the planetary temperatures across all longitudes and latitudes at several pressure layers. We note that utilizing a Gaussian in latitude is a simplification, and future work should utilize the results of GCMs to improve the latitude temperature formulation in a way similar to our longitudinal parameterization.

## 5. Summary

In this paper, we presented a novel physically motivated, parameterized multidimensional model for phase-curve retrieval that accounts for the advection of energy due to the planetary jet. This parameterization, informed by GCM models, accounts for the energy redistribution around the planet by including both radiative and advective processes. Instead of treating each phase independently, we fundamentally link them together and self-consistently redistribute energy throughout the entire planet via the equatorial jet. In addition to the resolved temperature structure, the model also returns essential information about the planetary jet structure, the location and extent of the jet, and its amplitude and phase offset. These recovered jet properties provide us with physical insight into the planetary energy budget and day–night redistribution, providing an important feedback for the GCMs.

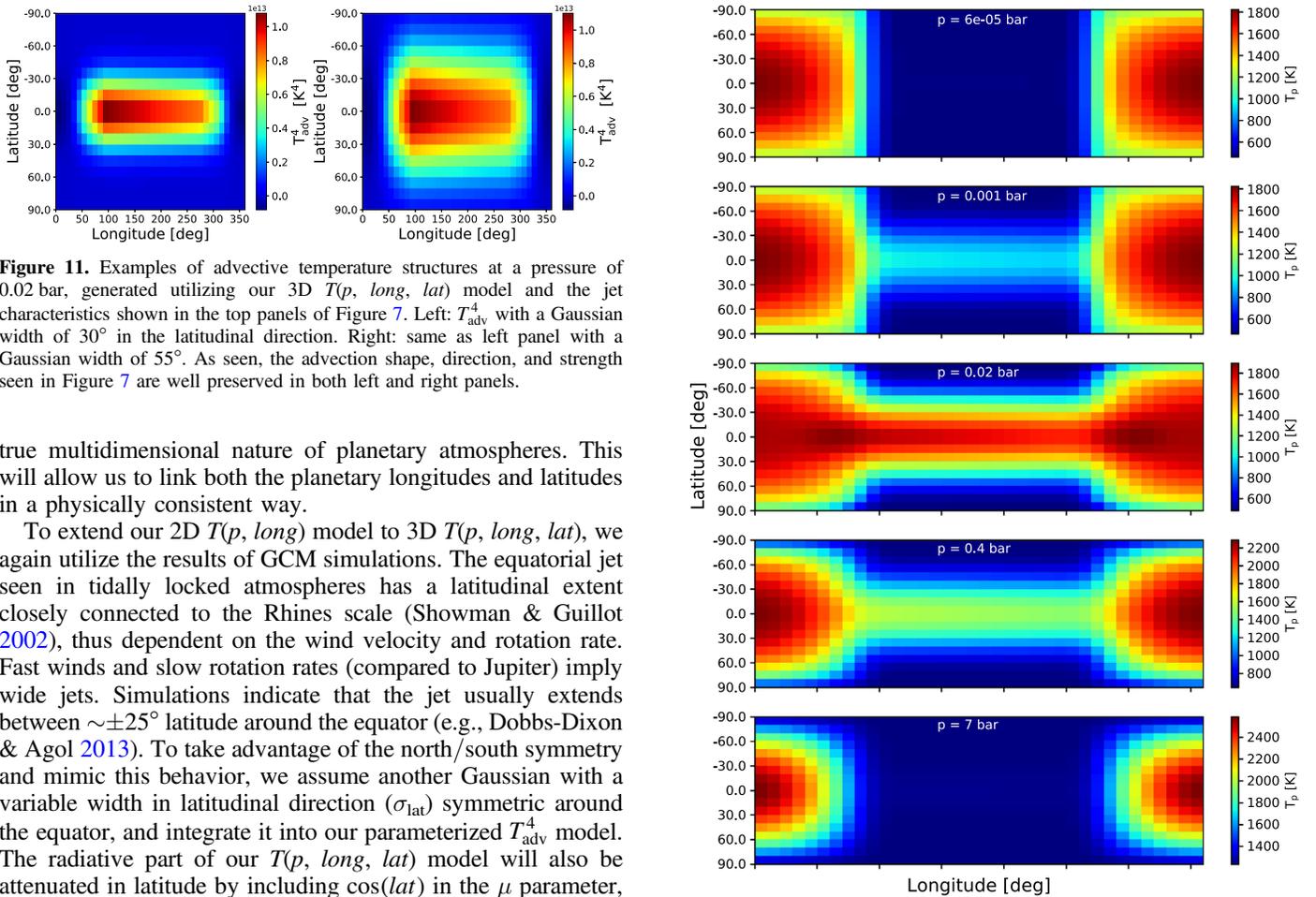

**Figure 12.** Planetary temperature across a range of pressures for the case shown in the top panel of Figure 11 and the jet characteristics shown in the top panel of Figure 7.

In this paper we presented only the forward model. In a following paper we will utilize this model in a simultaneous retrieval of the spectra from all orbital phases using phase-curve observations. Furthermore, in our approach we have concentrated on emission spectra, but this model can easily be generalized to incorporate transit spectra, where independent modifications to the temperature profiles at the eastern and western terminators can manifest themselves as changes in the spectra (Dobbs-Dixon et al. 2012).

Our parameterization is specifically targeted as a tool for retrieval analyses. The formulation is analytic and executes very quickly, returning the temperature structure of the entire planet in microseconds. The model utilizes an amenable set of physical parameters, efficient enough to enable Bayesian analysis, and returns physical insights not yet retrieved for exoplanets. In addition, it allows for the spectral retrieval of all planetary phases simultaneously. Since planetary jets in close-in, tidally locked planets have profound effects not only on the overall atmospheric temperature structure but also on the multidimensional clouds and chemistry, once this model is implemented in retrieval it can be further coupled with parametrized or self-consistent clouds and a prescription for equilibrium or nonequilibrium chemical transport and compositions, to allow for a thorough 3D characterization of exoplanetary atmospheres.

We thank the anonymous reviewer for insightful comments that improved the manuscript. I.D.D. and J.B. were partially supported by the NASA Exoplanets Research Program, grant No. NNX17AC03G. We thank Tristian Guillot for useful discussions. This work additionally benefited from the 2019 Exoplanet Summer Program in the Other Worlds Laboratory (OWL) at the University of California, Santa Cruz, a program funded by the Heising-Simons Foundation. Portions of research were carried out on the High Performance Computing resources at New York University Abu Dhabi.

## ORCID iDs

Ian Dobbs-Dixon  https://orcid.org/0000-0002-4989-6501
Jasmina Blecic  https://orcid.org/0000-0002-0769-9614